\documentclass[10pt,reqno]{amsart}
     \makeatletter
     \def\section{\@startsection{section}{1}%
     \z@{.7\linespacing\@plus\linespacing}{.5\linespacing}%
     {\bfseries
     \centering
     }}
     \def\@secnumfont{\bfseries}
     \makeatother
\theoremstyle{definition}

\theoremstyle{remark}

\numberwithin{equation}{section}
\begin{document}

\title[Wild Randomness, and Hyperbolic Diffusion in Financial Modelling]{Wild Randomness, and the application of Hyperbolic Diffusion in Financial Modelling}

\author{Will Hicks}
\address{Will Hicks: Memorial University of Newfoundland, St Johns, NL A1C 5S7, Canada}
\email{williamh@mun.ca}




\begin{abstract}
The application of the Cauchy distribution has often been discussed as a potential model of the financial markets. In particular the way in which single extreme, or ``Black Swan'', events can impact long term historical moments, is often cited. In this article we show how one can construct Martingale processes, which have marginal distributions that tend to the Cauchy distribution in the large volatility limit. This provides financial justification to the approach investigated in \cite{Romero}, and highlights an example of how quantum probability can be used to construct non-Gaussian Martingales. We go on to illustrate links with hyperbolic diffusion, and discuss the insight this provides.
\end{abstract}

\maketitle

\section{Introduction:}
In \cite{Mandelbrot}, Mandelbrot distinguishes between the concepts of `mild' randomness, for example that described by a normal distribution, which consists of fluctuations about a well defined mean, and `wild' randomness, for example that described by a Cauchy distribution. The Cauchy distribution has undefined moments, including an undefined mean, and describes a random variable whereby single extreme observations can have a significant effect on the the overall statistical properties.

Mandelbrot also presents evidence, in \cite{Mandelbrot}, that financial market returns are often better described by wild randomness, in the sense that `black swan' events occur from time to time that significantly distort the mean return observed over a historical period. This is despite the fact that, for reasons of pragmatism, models applied by practitioners in the financial services industry are invariably based on models of `mild' randomness.

The Black-Scholes model is derived using the principles of non-arbitrage, and through applying the Martingale Representation Theorem (see for example \cite{Oksendal}) to represent the financial market price as an Ito process.

In \cite{Romero} the authors attempt to generalise the Black-Scholes equation using relativistic quantum mechanics. The authors map the Klein-Gordon equation back to a new equation in the same way that the Schr{\"o}dinger equation is mapped to the conventional Black-Scholes (for example see \cite{Baaquie}). The key motivation for doing so, is to investigate the behaviour of the market in the large volatility limit.

Under these extreme conditions, the authors find that the processes driving the log returns for a stock price tend to a Cauchy distribution, and they note that this is in line with analysis carried out on the chaotic behaviour of the financial market by Mandelbrot in \cite{Mandelbrot}. Therefore, the motivation for studying such models is clear.

However, despite the appeal of the approach, there are issues to overcome. Generally, when deriving the partial differential equations of mathematical finance, one starts by expanding the function representing the derivative price ($u(x,t)$) to first order in $dt$ (see for example \cite{Kuo}):
\begin{align}\label{rel_Kolm}
dx=\sigma(x,t)dW\nonumber\\
du=\bigg(\frac{\partial u}{\partial t}+\frac{\sigma(x,t)^2}{2}\frac{\partial^2 u}{\partial x^2}\bigg)dt+\frac{\partial u}{\partial x}\sigma(x,t)dW
\end{align}
Then, taking expectations in the Martingale measure, we end up with a differential equation that is 2nd order with respect to the risky underlyings (eg the forward price: $x$) and first order with respect to time. Therefore, this approach cannot be used to justify the use of an equation such as the Klein-Gordon equation, which is second order with respect to time, for modelling in finance.

In this article we investigate these issues using quantum probability. First in section \ref{Q_app}, we discuss some of the principles in relation to this quantum approach. Then in section \ref{KGE}, we show how the pseudo-Hermiticity of the Klein-Gordon Hamiltonian (see \cite{Mostafazadeh}, \cite{Mostafazadeh2}) can be used to enforce the Martingale condition and the extent to which a non-arbitrage model can be constructed. Finally, in section \ref{hyper_diff}, we show how the resulting equation can be equated to hyperbolic diffusion equations, and discuss insights that can be obtained from this comparison.

\section{Background on the Quantum Approach:}\label{Q_app}
The application of quantum methods to mathematical finance has been investigated by a number of sources. One example approach is suggested by Accardi \& Boukas in \cite{AB}. The authors develop a quantum Black-Scholes equation using the methods of quantum stochastic calculus (see \cite{HP}), in the place of the standard Ito calculus. In this approach, the derivative price is modelled as an observable acting on the Market, which is represented by a state that exists within a Fock space. The authors develop operators that define the evolution of the market state, by adding noise to the Fock space. This noise is the equivalent of the usual Ito diffusion within a classical space.

In \cite{Hicks4}, the author discusses the nature of drift in such a quantum model. The related notion of the {\em closed} quantum Black-Scholes is suggested. In this model, it is found that, unlike randomness introduced from an external source, the intrinsic randomness associated to the quantum state does not necessarily introduce the usual property of strictly positive time value for the holder of an option contract (such as a vanilla call option contract).

In this article, we take a different approach. We model the price of a financial instrument using a wave-function that is described by a Schr{\"o}dinger equation. For example, if we wished to model a financial underlying that followed a Gaussian process, we could use the following equation:
\begin{align}\label{gaussian}
i\frac{\partial\psi}{\partial t}=\hat{H}\psi\\
\hat{H}=\frac{\partial^2}{\partial x^2}
\end{align}
Under this model, the probability density for the market price $x$, at time $t$, given the solution $\psi(x,t)$ to \ref{gaussian}, is given by:
\begin{align*}
p(x,t|\psi)=\langle\delta_x|\psi\rangle\\
=\psi(x,t)
\end{align*}
Here, $p(x,t|\psi)$ represents the probability density for measuring the price $x$ at time $t$, where the initial quantum state is given by: $\psi$. We write: $\delta_x$ for the generalized state such that the price is given by $x$, with probability $1$.

Under this interpretation, one can apply the mapping $it\rightarrow\tau$ (often referred to as a Wick rotation), under which \ref{gaussian} is mapped to the standard heat equation. This link has been investigated by a number of sources. For example, Haven discusses the implications of modelling a derivative price as the state function in a Schr{\"o}dinger equation in \cite{Haven}, \cite{Haven2}. In addition, the path integral approach to financial modelling has been investigated in \cite{Baaquie}-\cite{Baaquie3}, \cite{Linetsky}.

In some senses this approach is justified on mathematical rather than financial grounds. We are modelling using a closed system, and there is no external source of randomness. As discussed in \cite{Hicks4}, it is generally such external randomness that drives classical diffusion like properties and associated positive time-value for option contracts.  However, the approach allows us to define a probability measure with the desired statistical properties, and which can interpreted as a classical diffusion, via the Wick rotation.

\section{Financial Modelling using the Klein-Gordon Equation:}\label{KGE}
In relativistic classical mechanics, the energy of a free particle is specified by the energy-momentum 4-vector. With the speed of light, $c$, set to $1$, this is given by: $(E,p_x,p_y,p_z)$. The theory of special relativity implies that the scalar product (in the Minkowski metric) of this vector with itself gives the square of the rest mass: $E^2-|\mathbf{p}|^2=m^2$. Finally, by quantizing in the usual way: $E\rightarrow i\partial/\partial t$, and $p_x\rightarrow -i\partial/\partial x$, we get the Klein-Gordon equation (in one spatial dimension):
\begin{align}\label{KG}
\frac{\partial^2\psi}{\partial t^2}=\frac{\partial^2\psi}{\partial x^2}-m^2\psi
\end{align}
One can write \ref{KG} as follows (see \cite{Folland}, section 4.1):
\begin{align*}
\frac{\partial}{\partial t}\begin{pmatrix}\psi_1\\\psi_2\end{pmatrix}=\begin{pmatrix} 0 & 1\\ \nabla_x^2-m^2 & 0\end{pmatrix}\begin{pmatrix}\psi_1\\\psi_2\end{pmatrix}
\end{align*}
Since the matrix operator is not skew-adjoint, unlike the Schr{\"o}dinger equation, the Klein-Gordon equation is not Hermitian. Therefore, further work is required to provide the necessary risk neutral justification. We must first investigate the pseudo-Hermiticity of the Klein-Gordon equation.
\subsection{Pseudo Hermiticity of the Klein-Gordon Equation:}
A detailed study on the pseudo-Hermiticity of the Klein Gordon operator can be found at \cite{Mostafazadeh} and \cite{Mostafazadeh2}. In this section we simply quote the results we wish to apply in a financial modelling context.

As noted in \cite{Mostafazadeh} and \cite{Mostafazadeh2}, to produce a valid quantum framework, the Hamiltonian operator must conserve probability and have a real valued spectrum. This is guaranteed where the Hamiltonian is Hermitian. However, given a linear positive-definite automorphism: $\eta^+$, and an inner product defined by:
\begin{align*}
\langle\phi|\psi\rangle_{\eta^+}=\int_{\mathbb{R}}\overline{\phi(y)}(\eta^+\psi)(y)dy 
\end{align*}
then, a Hamiltonian operator: $\hat{H}$ will conserve probability and have a real valued spectrum in the event that it is {\em pseudo Hermitian} with respect to $\eta^+$. In other words, if:
\begin{align*}
\hat{H}^{\dagger}=\eta^+\hat{H}\eta^{+-1}
\end{align*}
To derive the required automorphism (and associated inner product), we follow \cite{Mostafazadeh2}. First write equation \ref{KG} as a Schr{\"o}dinger equation on the Hilbert space, $\mathcal{H}=L^2(\mathbb{R}^n)\oplus L^2(\mathbb{R}^n)$ (with arbitrary $\lambda\in\mathbb{R}-{0}$):
\begin{align}\label{KG2}
i\frac{\partial\Psi(t)}{\partial t}=\hat{H}\Psi(t)\\
\Psi(t)=\begin{pmatrix} \psi(t)+i\lambda\dot{\psi}(t)\\ \psi(t)-i\lambda\dot{\psi}(t)\end{pmatrix}\nonumber\\
\hat{H}=\frac{1}{2}\begin{pmatrix} \lambda D+\lambda^{-1} & \lambda D-\lambda^{-1}\\-\lambda D+\lambda^{-1} & -\lambda D-\lambda^{-1}\end{pmatrix}\nonumber
\end{align}
In \cite{Mostafazadeh} Mostafazadeh shows that \ref{KG2} is Pseudo-Hermitian, under the automorphism:
\begin{align}\label{KG_eta}
\eta^+=\frac{1}{8}\begin{pmatrix} \lambda^2+D^{-1} & \lambda^2-D^{-1}\\\lambda^2-D^{-1} & \lambda^2+D^{-1} \end{pmatrix}\\
D=-\frac{\partial^2}{\partial x^2}+\mu^2\nonumber
\end{align}
Since $\eta^+$ is positive definite, this implies that the Hamiltonian defined by \ref{KG2} has a real valued spectrum, and conserves probability under the positive definite inner product:
\begin{align}\label{KG_ip}
\Phi=\begin{pmatrix} \phi+i\lambda\dot{\phi}\\ \phi-i\lambda\dot{\phi}\end{pmatrix}\nonumber\\
\Psi=\begin{pmatrix} \psi+i\lambda\dot{\psi}\\ \psi-i\lambda\dot{\psi}\end{pmatrix}\nonumber\\
\langle\phi|\psi\rangle_{\eta^+}=\frac{1}{2}\Big(\langle\phi|\psi\rangle +\int_{\mathbb{R}} dx\int_{\mathbb{R}} dy\overline{\dot{\phi}(x)}G(x-y)\dot{\psi}(y)\Big)\nonumber\\
G(u)=exp(-\mu(|u|)/4\pi |u|
\end{align}
\subsection{Application in Finance: Using a Pseudo-Hermitian Hamiltonian}
If, instead of the relation given in \cite{Mostafazadeh}, we use:
\begin{align*}
\Psi(t)=\begin{pmatrix} e^{-it/\lambda}\psi(t)+i\lambda e^{-it/\lambda}\dot{\psi}(t)\\ e^{-it/\lambda}\psi(t)-i\lambda e^{-it/\lambda}\dot{\psi}(t)\end{pmatrix}
\end{align*}
the Schr{\"o}dinger equation defined by the new Hamiltonian (\ref{KG2}) is given by:
\begin{align*}
D\psi+i\lambda\dot{\psi}=-\frac{i\psi}{\lambda}+2i\dot{\psi}-\lambda\ddot{\psi}\\
-D\psi+i\lambda\dot{\psi}=-\frac{i\psi}{\lambda}+\lambda\ddot{\psi}
\end{align*}
This equation is equivalent to the following equation:
\begin{align*}
\lambda\frac{\partial^2\psi}{\partial t^2}-i\frac{\partial\psi}{\partial t}+D\psi=0\\
\end{align*}
If we set: $D=-\frac{\sigma^2}{2}\frac{\partial^2}{\partial x^2}+\mu^2$, then we get:
\begin{align}\label{Sch_eqn}
i\frac{\partial\psi}{\partial t}-\lambda\frac{\partial^2\psi}{\partial t^2}=-\frac{\sigma^2}{2}\frac{\partial^2\psi}{\partial x^2}+\mu^2\psi
\end{align}
In the next sections, we review the following:
\begin{itemize}
\item The limiting case $\lambda\rightarrow 0$, section \ref{BS_Case}.
\item The Cauchy distributed solution in the case $\mu\rightarrow 0$, section \ref{Cauchy_case}.
\item Enforcing the Martingale condition in section \ref{Mart_Cond}.
\item Links to the hyperbolic heat conduction equation, section \ref{hyper_diff}.
\end{itemize}
\subsection{The Limiting Case $\lambda\rightarrow 0$.}\label{BS_Case}
First define:
\begin{equation*}
L_{\lambda}\psi=\lambda\frac{\partial^2\psi}{\partial t^2}-\frac{\sigma^2}{2}\frac{\partial^2\psi}{\partial x^2}+\mu^2\psi
\end{equation*}
So that we have:
\begin{equation*}
i\frac{\partial\psi}{\partial t}=L_{\lambda}\psi
\end{equation*}
Furthermore, we denote:
\begin{equation}
\hat{H}\psi=-\frac{\sigma^2}{2}\frac{\partial^2\psi}{\partial x^2}+\mu^2\psi
\end{equation}
If we use the norm defined by: $||\psi||=\sqrt{\langle\psi|\psi\rangle}$, then we have:
\begin{equation*}
||L_{\lambda}\psi-\hat{H}\psi||=\lambda\bigg|\bigg|\frac{\partial^2\psi}{\partial t^2}\bigg|\bigg|
\end{equation*}
Now strictly speaking, since $L_{\lambda}$ and $\hat{H}$ are unbounded operators they can only be defined on a dense subset of the Hilbert space $L^2(\mathbb{R})$. However, assuming $\psi$ is in the intersection of the domains of $\hat{H}$, $L_{\lambda}$ and $\partial^2/\partial t^2$, then we have, for $\lambda\rightarrow 0$:
\begin{equation*}
||L_{\lambda}\psi-\hat{H}\psi||\rightarrow 0
\end{equation*}
Now it follows from the analysis given in \cite{Jana}, that the Black-Scholes model can be defined using the Hamiltonian:
\begin{align*}
\hat{H}_{BS}\psi=-\frac{\sigma^2}{2}\frac{\partial^2\psi}{\partial x^2}+\frac{\sigma^2}{8}\psi
\end{align*}
Therefore, we find that in the limiting case of $\lambda\rightarrow 0$, if $\mu^2\rightarrow \sigma^2/8$, then the random variable will match the quantum version of the arbitrage free Black-Scholes equation:
\begin{align*}
i\frac{\partial\psi}{\partial t}=\hat{H}_{BS}\psi
\end{align*}
\subsection{The Cauchy Distributed Solution.}\label{Cauchy_case}
Now let $\psi$ be a solution to equation \ref{Sch_eqn}. We have seen that for small $\lambda$, the solution will be similar to the Black-Scholes model. In this section we consider the model, in the limit $\mu^2\rightarrow 0$, and for large $\lambda$.

Let $\psi$ be a solution to equation \ref{Sch_eqn}. Further, let $f(t)$ represent a differentiable complex function of time. Then we have:
\begin{align*}
\frac{\partial^2}{\partial t^2}(f(t)\psi)=\frac{\partial^2 f}{\partial t^2}\psi+2\frac{\partial f}{\partial t}\frac{\partial\psi}{\partial t}+f(t)\frac{\partial^2\psi}{\partial t^2}\\
\frac{\partial^2}{\partial x^2}(f(t)\psi)=f(t)\frac{\partial^2\psi}{\partial x^2}
\end{align*}
Therefore, if we set $f(t)=exp(-it/2\lambda)$, then we can write equation \ref{Sch_eqn} as:
\begin{align}
\frac{\lambda}{f(t)}\frac{\partial^2(f(t)\psi)}{\partial t^2}=-\frac{1}{f(t)}D(f(t)\psi)\nonumber\\
D=-\frac{\sigma^2}{2}\frac{\partial^2}{\partial x^2}+\bigg(\mu^2+\frac{1}{4\lambda}\bigg)
\end{align}
After carrying out the Wick rotation $\tau=-it$, taking the limit $\mu^2\rightarrow 0$, and multiplying both sides by $f(t)$, we get:
\begin{equation}\label{Cauchy}
\lambda\frac{\partial^2(f(\tau)\psi)}{\partial \tau^2}+\frac{\sigma^2}{2}\frac{\partial^2(f(\tau)\psi)}{\partial x^2}=\bigg(\frac{1}{4\lambda}\bigg)f(\tau)\psi
\end{equation}
Next we make the substition $z=x\sqrt{2\lambda/\sigma^2}$ to get:
\begin{equation}\label{Cauchy2}
\frac{\partial^2(f(\tau)\psi)}{\partial \tau^2}+\frac{\partial^2(f(\tau)\psi)}{\partial z^2}=\bigg(\frac{1}{4\lambda^2}\bigg)f(\tau)\psi
\end{equation}
As shown in \cite{Romero}, the equation:
\begin{equation*}
\frac{\partial^2(f(\tau)\psi)}{\partial \tau^2}+\frac{\partial^2(f(\tau)\psi)}{\partial z^2}=0
\end{equation*}
Has a Cauchy distributed solution:
\begin{equation*}
f(\tau)\psi_C(z,\tau)=\frac{1}{\pi}\frac{\tau}{z^2+\tau^2}
\end{equation*}
Feeding this back into equation \ref{Cauchy2}, we find:
\begin{itemize}
\item $f(\tau)\rightarrow 1$ as $\lambda\rightarrow\infty$.
\item The right hand side of \ref{Cauchy} reduces with $O(\lambda^{-2})$ as $\lambda\rightarrow\infty$.
\item The coefficients on the left hand side are independent of $\lambda$
\end{itemize}
Therefore, in the limit of large $\lambda$, we find our solution tends to a Cauchy distribution.
\subsection{The Martingale Condition:}\label{Mart_Cond}
We assume that the Martingale condition can be written:
\begin{equation}\label{Mart_Cond_eqn}
E^{\psi}[X|x_0]=x_0
\end{equation}
Where, $x_0$ represents the current observed market price ($t=0$). Then, we take expectation with respect to the probability density function $\psi$.

The solution to \ref{KG2} yields two independent functions: $\psi(x,t)$, and $\dot{\psi}(x,t)$. As noted in \cite{Mostafazadeh} inner product \ref{KG_ip} is both positive definite, and time invariant. It can therefore be used to construct a probability interpretation. To do this, using \ref{KG_ip}, we assume the probability density function, conditional on the value at $t=0$, is given by:
\begin{align*}
p(x,t|x_0,0)=\frac{1}{2}|\psi(x,t)|^2+\frac{1}{2}\bigg\rvert\dot{\psi}(x,t)\int_{\mathbb{R}}dyG(x-y)\dot{\psi}(y,t)\bigg\lvert\\
\Psi(x,t)=\begin{pmatrix} \psi(x,t)+i\lambda\dot{\psi}(x,t)\\ \psi(x,t)-i\lambda\dot{\psi}(x,t)\end{pmatrix}
\end{align*}
So finally, the Martingale condition can be expressed in this case as:
\begin{align}
x_0=\frac{1}{2}\int_{\mathbb{R}}x|\psi(x,t)|^2dx+\frac{1}{2}\int_{\mathbb{R}}x\bigg\rvert\dot{\psi}(x,t)\int_{\mathbb{R}}dyG(x-y)\dot{\psi}(y,t)\bigg\lvert dx
\end{align}
\section{Insights from The Hyperbolic Diffusion Equation:}\label{hyper_diff}
In general, classical diffusion models can be obtained by combining the equations:
\begin{align}
\frac{\partial u}{\partial t}=\nabla\cdot\mathbf{q}\label{cont}\\
\mathbf{q}=-K\nabla u\label{flux}
\end{align}
Let $u(x,t)$ represent the number of particles at time $t$ in an infinitesimally small region around $x$, and $\mathbf{q}$ represent the flux of the particle flow. From a financial perspective, one can interpret individual particles as a discrete unit of probability. For example, each the path of each ``particle'' can represent a single Monte-Carlo sample path, in a simulation of the financial market.

Then equation \ref{cont} simply represents the continuity equation: the number of particles in the system is constant. The equivalent financial analogy is that the total probability in the system must always add to one.

Equation \ref{flux}, represents the fact that the movement of particles is purely random, and that an increased concentration of particles in one region will result in more particles randomly flowing out of the region, than randomly flowing in. In other words, whilst the total probability in the system as a whole must add to 1, there will in general be a build up of probability in some price ranges, and a reduction in probability in others.

In \cite{Gomez}, the authors show that by amending equation \ref{flux} to:
\begin{equation}\label{flux2}
\mathbf{q}+\lambda\frac{\partial\mathbf{q}}{\partial t}=-K\nabla u
\end{equation}
one generates the hyperbolic diffusion equation:
\begin{align}\label{HHCE}
\lambda\frac{\partial^2 u}{\partial t^2}+\frac{\partial u}{\partial t}=K\frac{\partial^2 u}{\partial x^2}
\end{align}
Similarly, if one applies a Wick rotation; $\tau=it$ to equation \ref{Sch_eqn} with $\mu^2=0$, one obtains:
\begin{align}\label{BackwardHHCE}
\frac{\partial u}{\partial \tau}+\lambda\frac{\partial^2 u}{\partial\tau^2}=\frac{\sigma^2}{2}\frac{\partial^2 u}{\partial x^2}
\end{align}
So we can see that, in the same way that the Gaussian solutions of the conventional heat equation are linked to the Schr{\"o}dinger equation with Hamiltonian:
\begin{align*}
\hat{H}=-\frac{\sigma^2}{2}\frac{\partial^2}{\partial x^2}
\end{align*}
The solutions in this case, with $\mu^2=0$, are linked to the Hyperbolic Heat equation.
\subsection{Interpretation of the Flux Equation:}
In this model, due to equation \ref{flux2}, a gradient in the concentration of particles can coexist with zero flux, when it drives instead an {\em acceleration} in the flux. In other words the flow of particles doesn't react instantaneously to an asymmetric distribution of particles, but reacts over a time period characterised by $\lambda$. A small value for $\lambda$, indicates that the flux will accelerate quickly.

In equation \ref{flux}, since the diffusive step is purely random. Diffusing one way is no more or less likely than another. The flux is simply determined by the gradient of the probability density. Particles diffuse from high density zones to lower density zones simply due to the balance of probability.

For equation \ref{flux2} in the large $\lambda$ limit, $\sqrt{K/\lambda}$ acts as the propagation speed in the associated wave equation. From a Newtonian wave-mechanics perspective, a non-zero gradient in the probability density actually drives particles apart. Since the left hand size of \ref{flux2} represents an acceleration, the right hand side can be interpreted as a Newtonian force. Particles that are in high density zones are pushed away into lower density zones. One could imagine springs between particles, whereby the mass is given by $\lambda$, and the spring constant by $K$.

The financial interpretation of increasing $\lambda$ is probability density functions, that have increased propensity to spread out along the price axis.
\subsection{Explicit Finite Difference:}
An explicit finite difference method shows that the vector at time: $n+1$, is given by (where $L$ is the discretized second derivative operator):
\begin{align}\label{FD}
\mathbf{u}^{n+1}=\mathbf{u}^{n}+\bigg(\delta t-\frac{\delta t^2}{2\lambda}\bigg)\frac{\partial\mathbf{u}^n}{\partial t}+\bigg(\frac{K\delta t^2}{2\lambda}\bigg)L\mathbf{u}^n
\end{align}
From this it is clear that:
\begin{itemize}
\item From the discretization \ref{FD}, it is clear, that the next diffusive time-step is driven both by the current probability density/concentration of particles ($\mathbf{u}^n$) {\em and} its current rate of change ($\partial\mathbf{u}^n/\partial t$). This can also be applied in the financial context. The dynamics of a market in a rapidly changing state are likely to be very different to the market that has been stable over recent history.
\item This conclusion is consistent with the quantum approach provided above, where the general solution to equation \ref{Sch_eqn} involved 2 independent functions $\psi(x,t)$, and $\dot{\psi}(x,t)$.
\item At first glance the dependence of the diffusion on the initially observed rate of change destroys the possibility of using this approach in a non-arbitrage pricing framework. However, there are many real properties of the financial markets, such as a variance that doesn't scale linearly with time, that do not fit the stringent requirements for a non-arbitrage approach. Therefore, we leave open the question of how the relativistic quantum Black-Scholes can be applied in practice.
\item Lastly, if one wishes to model in the large $\lambda$ limit, one must scale the diffusivity $K$ (or $\sigma^2$) accordingly. 
\end{itemize}
\section{Conclusions:}
In this article we have sought to provide justification to the approach suggested in \cite{Romero}. The result is a model, that in the small $\lambda$ limit, behaves like a standard Black-Scholes model, and in the large $\lambda$ limit tends to models of wild randomness, such as the Cauchy distribution.

We have also shown how links to the hyperbolic heat equation can provide some understanding regarding the nature of the dynamics of the models, and also numerical methods that may be applied.

Further work is required to incorporate the model into a real model of the financial market, and to decide the extent it can be embedded within a no arbitrage pricing framework.
\bibliographystyle{amsplain}

\end{document}